# On Policy Recommendations from Causal Inference in Physics Education Research

M. B. Weissman

*Department of Physics, University of Illinois at Urbana-Champaign*

*1110 West Green Street, Urbana, IL 61801-3080*

**Abstract**

Sound educational policy recommendations require valid estimates of causal effects, but observational studies in physics education research sometimes have loosely specified causal hypotheses. The connections between the observational data and the explicit or implicit causal conclusions are sometimes misstated. The link between the causal conclusions reached and the policy recommendations made is also sometimes loose. Causal graphs are used to illustrate these issues in several papers from Physical Review Physics Education Research. For example, the core causal conclusion of one paper rests entirely on the choice of a causal direction although an unstated plausible alternative gives an exactly equal fit to the data.

## Introduction

A central goal of most physics education research (PER) is to find ways teach better, i.e. to improve educational outcomes. That requires determining the probable results of several possible courses of action, so that value judgments can be applied to choose the best policies. Problems of the form "If we do this vs. that, what will be the probable difference in results?" are by definition problems in counterfactual or potential outcome causation. (See (*1*) for a standard text, or (*2*) for an up-to date technical review.) Techniques for drawing causal conclusions from randomized controlled trials are generally understood, but the traditional methods for drawing causal inferences from observational data (e.g. multiple regression) are often inappropriate or misapplied. (*1, 2*) As PER has moved to tackle increasingly challenging problems, the causal inference challenges have grown. "Correlation does not imply causation" is a truism that is easier to state than to follow in practice.

The purpose of this paper is to illustrate the need for better causal analysis in PER by using a few instances of papers published in Physical Review Physics Education Research (PRPER) in which questionable procedures were used either to draw causal inferences from data or to draw policy recommendations from causal inferences or both. It is motivated by exchanges both with statisticians discontented with the level of causal reasoning in many social sciences and with colleagues in PER who were both uneasy about their lack of familiarity with modern techniques and curious about how well some recent PER work would hold up under scrutiny. I've chosen several papers by leaders in PER to illustrate what are said to be some common issues

I call attention to these issues because improvement is possible. In recent decades more valid methods of causal inference have been developed (*1, 2*) with applications in epidemiology(*2*), biology(*3*), public health(*4*), economics(*5*), psychology(*6*), sociology(*7*), political science(*8*), education(*9*), and other fields. Although here I will focus on instances in which causal conclusions rely on unstated questionable assumptions, the new methods allow causal conclusions to be made from complicated data so long as reasonable explicit assumptions are made. (*1, 2*)

I will use only the elementary level of analysis with which I'm familiar. Other than encouraging more study of causal inference, I will make no educational recommendations, but will suggest an editorial change for the field.

Before presenting an admittedly blunt critique of the causal reasoning in these PER papers, I should acknowledge that the much of the correlational data and analysis presented in them may well prove useful. My point is that invalid causal conclusions can be found even in papers treating correlational results carefully. Some additional data, ideally obtained by measuring effects of various interventions, will be needed in order to draw reliable causal conclusions to be used in making plans. . (A version of this paper is currently in press in PRPER, but this arXiv paper includes discussion of two papers(*10*) (*11*) that were published after my PRPER paper was submitted, so those sections have not been through the peer-review process.)

## **Background on Causal Methods**

Graphical representations of causal patterns(*1*), developed especially by Pearl and coworkers, are perhaps the most accessible of the newer methods and will suffice for this paper. Graphs are used to represent the most important feature of causal relations, that changing some quantities leads to changes in only some others. Graphs called Directed Acyclic Graphs (DAGs) represent this one-way feature by using arrows to show the causal directions. (*1*) A change in any variable only affects variables downstream from it in the DAG, and there are no causal loops. For physicists, the most familiar such relation is that changes propagate only forward in time, at least on the classical macro scale. A DAG itself does not give the functional form, magnitude, or even sign of the causal effects, but it does encode conditional independence relations, i.e. which interventions can have no effect on which other variables. DAGs are particularly useful for visualizing which variables should be controlled for, which should not be controlled for, and which create less tractable problems when estimating causal relations from correlations. (*1*)

A bidirectional arrow is sometimes used as a shorthand representation of an unmeasured variable with no arrows going in and arrows going out to two measured variables in a DAG. Graphs using that shorthand are called Acyclic Directed Mixed Graphs (ADMGs). Their bidirectional arrows should not be mistaken for bidirectional causation.

The papers to be discussed here assume that the relations among the variables can be well-approximated as linear. Then a linear Structural Equation Model (SEM) can explain the correlations among the variables in terms of the causal structure expressed in the ADMG.(*12*) Each causal arrow gets a coefficient showing the *direct* effect of changes of the tail variable on the head variable. The *net* causal effect of one variable on another is obtained by summing the products of the coefficients along each path leading from the first variable to the second.(*13*)

The directions of the arrows provide essential information needed to determine policy choices, since making a change in some quantity only affects variables downstream.(*1*) You can make a dog wag its tail by giving it a treat, but you cannot put a treat in its mouth by wagging its tail. Although words like "effect" and "impact" are often used to describe correlations while suggesting causation, the statements "Giving treats has a tail-wagging effect" and "tail-wagging has a treat-giving effect" do not mean the same thing. The difference would appear in the direction of the arrow between "treat" and "wag". The choice of graphs has major practical importance because they capture this causal distinction.

If the relevant ADMG is not already known, SEMs can only play a limited role in inferring causation from correlations. (*13*) Correlational data can rule out some ADMGs, e.g. ones with no pathways connecting variables that are significantly correlated, but that is far short of determining which ADMG should be used as the framework for the SEM. Often SEMs based on many different graphs can approximately fit the same data set. Furthermore, there are theorems showing that the ADMGs can be divided into Markov equivalence classes, each of whose members fit any given data set *exactly* equally well, with the same number of adjustable parameters. (*1-3, 14*) There are known rules for generating equivalence classes of graphs. (*1, 14*) On their own, SEMs provide evidence for choosing among equivalence classes but not for choosing graphs within a class. The choice among such equivalent graphs should be left open unless there is other evidence such as time order, other types of data, or firm knowledge about mechanisms, although intuition may guide tentative beliefs.(*13*)

In the first three papers described, particular graphs are chosen as the basis for SEMs but without sufficient reason to exclude other graphs with very different causal implications. I will use alternative members of equivalence classes to illustrate rigorously how big those causal differences can be, but not to advocate any particular causal model. A fourth paper uses SEMs carefully, but then draws strong causal conclusions about unmeasured variables outside the range described by its graph. A fifth paper, specifically intended to present methods, incorrectly describes how to find causal patterns from correlations and uses specific examples in which the causal questions are ill-defined and the methods inappropriate.

**Correlations do not determine causal graphs**

I shall discuss three papers (*10, 15, 16*) that say little about how the directions of graphical arrows were chosen, and whose causal conclusions turn out not to be justified by their data. These papers use SEMs as if they were a nearly stand-alone method for discovering causal patterns. (*10, 15, 16*)

Two of these papers (*15, 16*) look at traits like "Interest" rather than events. Gradually developed traits can be unrolled into functions of time. It is then inappropriate to represent a snapshot in time of a set of such traits by *any* ADMG since effects generally have been flowing *both ways* between past values of the traits. (*2*) (Fortunately, methods have been developed for inferring causal relations in complicated time-dependent longitudinal data. (*2, 17*)) Thus when I suggest alternatives to the causal interpretations used in these papers, I do not mean to imply that these alternatives are better and certainly not that they are right, since it is unlikely that any ADMG correctly represents the causal relations between snapshots of prolonged traits. Although this is a fundamental problem for the sorts of SEM models used in those papers, I will confine the discussion to narrower methodology issues that are relevant to a broader range of papers.

a) Gender

One paper(*15*) relates gender to questionnaire-based measures of attitudes toward physics, "…to explore which motivational factors cause changes in other factors…".(*15*) The attitudes are grouped into four clusters: "Interest", "Perceived Recognition", "Competency Belief", and "Physics Person", with the latter three abbreviated here as Recognition, Competence, and Identity for notation compatible with (*16* ). The linear SEM model used (*15*) is based on the DAG shown in Fig. 1 here. Since gender almost always precedes the attitudes examined, it can be assumed to have no incoming arrows. (Selection effects on who gets into the sample of calculus-based physics students could complicate that relation(*18*), but I shall not explore such effects here.)

This graph was chosen by "dropping connections or variables of low strength" from the saturated model in which all variables are connected, although the minimum strength for inclusion is not specified. (*15*) The two connections dropped were from Gender to Competence and to Identity. All six connections between the four core attitudinal variables are retained.

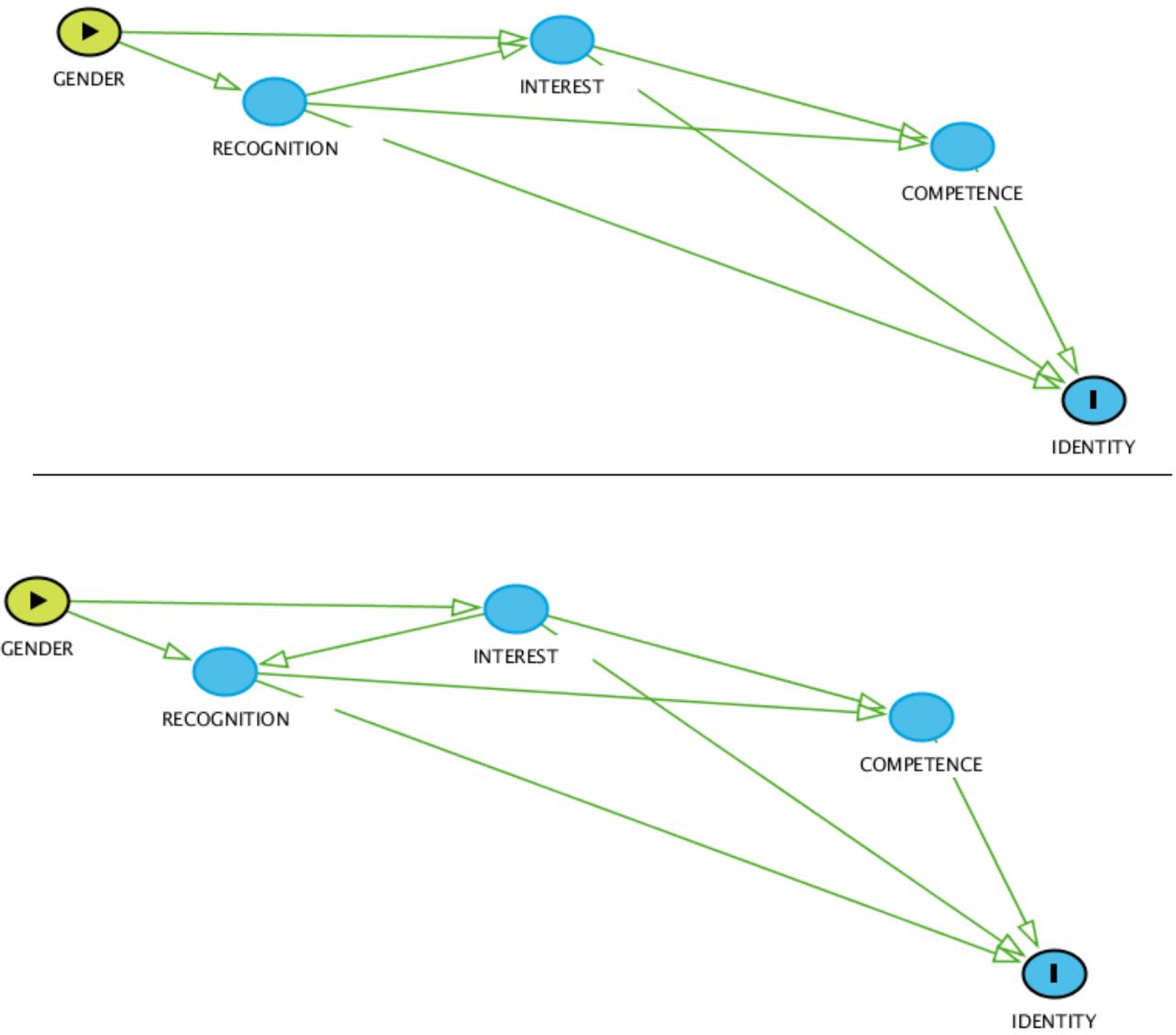

**Figure 1**. The top graph is the DAG from reference (*15*), using abbreviated variable names. The bottom one is the Markov equivalent graph described in the text. The graphs were drawn using the online DAGitty tool.

How the arrow *directions* were chosen is described only by: "…we only used the suggestions that were theoretically plausible". (*15*) Markov equivalence rules say the arrows between Interest and Recognition and between Identity and Competence can each independently be assigned either direction, leaving four equivalent DAGs. (*1, 14*) (Either of these two arrows could also be replaced by a bidirectional arrow giving nine equivalent ADMGs. (*14*))

Theoretical plausibility seems inadequate for deciding the directions of the arrows, especially between Interest and Recognition. Do we really know that students become interested in areas for which they perceive recognition more than that they perceive recognition for areas in which they have been interested? The equivalent graph shown in Fig. 1 obtained if the arrow from Recognition to Interest were reversed would have substantially changed meaning. The coefficient between Interest and Recognition would increase slightly, from 0.64 to 0.67. The coefficient from Gender to Interest would increase from 0.16 to 0.33, and the coefficient from Gender to Recognition would fall from 0.27 to 0.05. The qualitative impression from this equivalent graph would then be that gender differences in interests are most important, rather than gender differences in perceived recognition. The data do not provide guidance on which choice is better.

The very small coefficient for the Gender→Recognition arrow in this equivalent graph raises the question of whether it would meet the unspecified criteria for "dropping connections or variables of low strength", since it is only one third as large as the smallest coefficient included. (*15*) Based on the sample size of 559, (*15*) the 95% confidence interval for this coefficient would be approximately (-0.05, 0.15), so that conventionally one would expect the coefficient to be dropped from the model as statistically insignificant. If it were dropped, then in the resulting new equivalence class all Gender effects would be mediated through Interest. This equivalence class also includes some DAGs for which Recognition has no causal effects on other traits.

The claim of the title of ref. (*15*), that females "do not identify with physics [because] they do not think others see them that way" thus lacks support in the data. The data shown are *exactly* as compatible with a very small and statistically insignificant effect of gender on perceived recognition as they are with the moderate-size effect presented. Interpretations in which the Gender→Recognition and Recognition→ Identity effects are both zero are disfavored only by that insignificant coefficient. One simply can't tell from these data to what extent the paper's causal claims are correct.

The broad policy recommendations, centered on making the physics classroom gestalt more supportive, did not require careful identification of causal relations among the attitudinal variables. (*15*) An overlapping group of authors did report that a controlled intervention(*19*) encouraging effort when faced with difficulty and facilitating supportive interactions among students gave important improvements in physics course performance for females and perhaps also for non-white students, although the effect didn't reach conventional statistical significance for the small non-white sample.(*19*). It did not include a survey of the traits studied in (*15*) and (*16*), so it is hard to tell which, if any, of these traits were most affected. This experiment serves as a reminder of how much more straightforward it is to find out what works from interventions than from static correlations.

b) Out-of-class activities.
The effects of out-class science and engineering activities (OCSE) on student attitudes were explored in a recent paper for which "…the primary goal of the current analysis is determining the impact of OCSE activities on physics identity", i.e. estimating a causal effect. (*16*) The abstract raises a policy question based on causal effects: "Understanding the influence of students' science and engineering experiences on career choices is critical in order to improve future efforts…"(*16*)  The abstract also reaches a causal conclusion "out-of-class science and engineering activities have the largest influence on physics performance/competence beliefs…" (*16*) Although at points non-causal associative wording is used, the body of the paper is peppered with causal conclusions, e.g. "Recognition beliefs, while having the largest impact on overall identity…" and "…physics identities have less impact on their career choice… ". (*16*) In one case a conclusion is drawn explicitly about the expectation of what would happen if something is done:  "…if performance/competence beliefs are developed in isolation from recognition beliefs and interest, a student is not more likely to develop an overall physics identity." (*16*)

The causal conclusions are based on coefficients in an SEM relating OCSE to the same questionnaire-based measures of attitudes toward physics used in ref. (*15*), here called Interest, Recognition, Performance/Competence, and Physics Identity. (I again use Competence and Identity as abbreviations for the latter two.) The ADMG connecting these variables (*16*) is redrawn in DAG form as Fig. 2 below after converting the double-arrow of the original to an explicit unmeasured variable U.

Ref. (*16*) does not mention any criteria used in its choice of graphs, but says that the graph is based on prior work(*20*). That work says "we need to caution that the word effect in the following exploration of the results cannot imply causality in a strict sense…" and "… it is important to keep in mind that this study is correlational, so causality is not certain." The prior work also does not explore the class of graphs that are equivalent to the one they chose, (*20*) which include the same nine ADMGs connecting Interest, Recognition, and Competence we saw for ref. (*15*).

The graph chosen in ref. (*16*) modifies the one in (*20*) by adding OCSE, with three arrows coming out and none going in. (*16*) Those directions may not represent how the world works. If interested students often choose to do extra science activities, for example, Interest may have more causal effect on OCSE than OCSE has on Interest.

Nevertheless, if the causal assumptions implicit in the graph chosen are correct, then the "primary goal", "determining the impact of OCSE activities on physics identity" (*16*), is simply given by the unconditional regression coefficient of Identity on OCSE, since no confounders are present. (*1*) Neither this simple relation nor the resulting causal coefficient are given in the results. (*16*)

The quality of the SEM fit to the data is discussed but not in comparison to fits using other graphs. (*16*) The graph choice could not have been based on some standard trade-off between fit quality and number of parameters, e.g. the Akaike Information Criterion(*21*), because elementary DAG rules(*1*) say that reversing the arrow from Competence to OCSE gives a graph in the same equivalence class. That reversal, however, would make the unconditional regression of Identity on OCSE *not* equal to the impact of OCSE activities on physics identity and thus would change the implied estimate of "…the primary goal …the impact of OCSE activities on physics identity". There may also be other equivalence classes that fit better. (*16*)

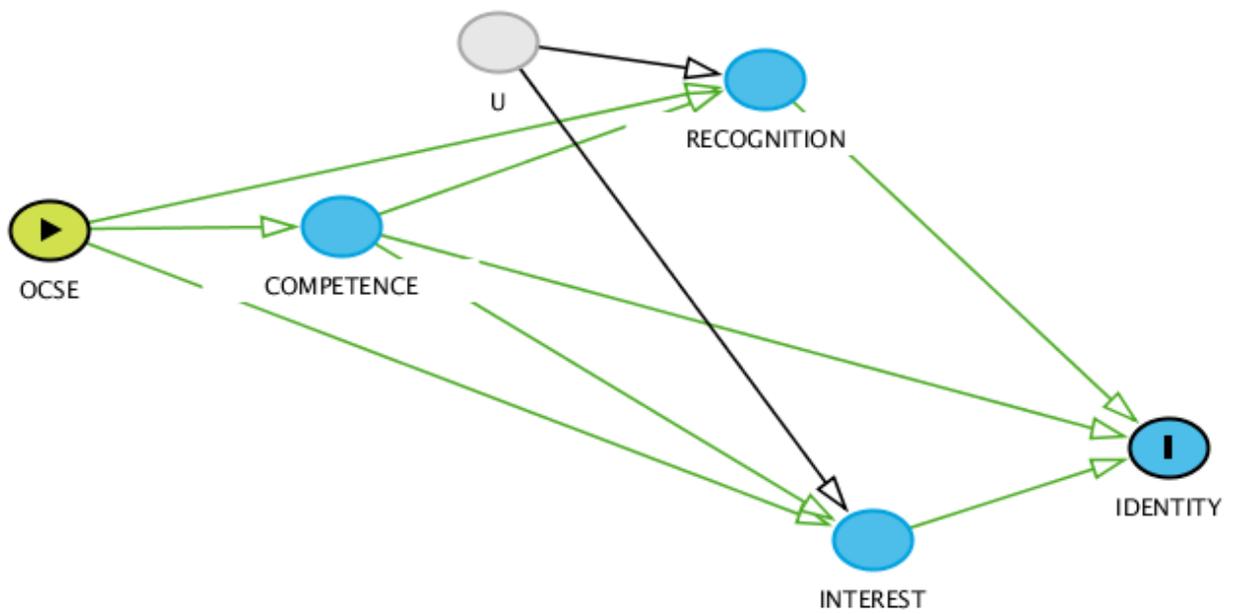

Figure 2. This is the DAG form of the ADMG used in reference (*16*), with their bidirectional arrow shown instead as an explicit unmeasured U. The variable "Physics Career Choice", connected only by an arrow from Identity, is omitted for simplicity.

The ADMG bidirectional arrow here has sufficient properties to allow replacement with either unidirectional arrow while preserving equivalence(*14*), giving many graphs in the same equivalence class. Even if we leave Identity fixed as a child of all but OCSE, there are 24 DAGs obtained from the permutations of the other labels in the saturated graph of the other four variables. (In addition there are many equivalent ways of substituting one or two bidirectional arrows for unidirectional ones.) Two of the equivalent DAGs would have reversed directions of all three arrows from OCSE. For those equivalent graphs "the impact of OCSE activities on physics identity" would be identically zero.

Despite the plethora of equivalent graphs available, there is one indication that the ADMG chosen does not represent causality very well. The authors remark upon the "surprising" negative sign of the regression coefficient for the direct effect of Competence on Identity. (*16*) That interpretation results from inappropriately treating the negative SEM coefficient found in a causally unmotivated graph as if it were a causal coefficient describing what would happen "…if performance/competence beliefs are developed in isolation from recognition beliefs and interest….". As an association, however, the negative coefficient is not surprising. A student who is confident in their abilities but still neither has interest in a field nor has sought recognition for it probably finds that field unattractive for some other reason and thus does not identify with it.

The "surprising" negative coefficient provides an example in which SEM results can help guide the formulation of causal hypotheses using graphical methods. This verbal description could be translated to a graph description by a slight modification of the graph used, which already includes an unmeasured cause U affecting both Interest and Recognition. U could be any number of causes that would be likely to have the same sign of effect on Identity as on Interest and Recognition. If there should also be a causal arrow from U to Identity with the same sign of coefficient as the arrows from U to Interest and Recognition, the regression coefficient obtained for Identity on Competence in the graph without the U→Identity arrow would differ from the direct causal coefficient by an undetermined amount of negative collider stratification bias, since Interest and Recognition are colliders between U and Competence.(*22*).

What sort of traits could contribute to the implicit U? Many factors are believed to affect how attractive a scientific field is to some student.(*23*) Anecdotally, it is easy to think of many friends who were good or even brilliant physics students but who knew early on that their strongest

8/7/21                                                                                                                                                      12

interests lay elsewhere. One previous study found a negative correlation between desiring a career with interpersonal interactions and identifying as a Physics Person(*24*). An experimental study from an overlapping group showed that discussing the underrepresentation of women in physics leads to more of them expressing interest in physics careers.(*25*) Closer integration of such experimental results with attitudinal survey data might produce clearer and more useful predictions. These informal suggestions are just meant as a reminder that a combination of explicit prior ideas about causality together with SEM results can suggest causal models with testable predictions. (*13*)

The paper's Conclusion draws policy recommendations based on what the effects would be of "Modifying programmed activities to better support recognition beliefs and interest…" or "…recognizing students in programmed OCSE activities … ". (*16*) The recommendation to "…[ensure] that activities are not only fun and engaging but also provide challenge. Students need to be provided with sufficient guiding support…" is not particularly reliant on the causal analysis. (*16*)

c) Comparing (a) and (b)

Both studies (*15*) (*16*) include all six connections among the four shared core attitudinal variables. One connection is represented with a unidirectional arrow in one study(*15*) and with a bidirectional arrow representing the influence of an implicit unmeasured variable in the other study(*16*). Two connections are represented with arrows pointing opposite directions in the two studies. Three are represented with arrows in the same direction in both studies. One of those three arrows has opposite signs of coefficients in the two studies. Of the six connections only two are represented with arrows of the same type with the same sign of coefficient. *It would be hard to argue that the two graphs representing these variables are close to converging toward a shared causal picture.*

No amount of linear SEM analysis can decide which causal pattern is best without further inputs. One obvious type of input would be longitudinal data. It may be that no ADMG could represent the causal relations among the traits unless each trait were unrolled into a time series. Specific interventions in randomized controlled trials would provide stronger evidence. For example, ref.(*26*) showed that an intervention designed to increase interest in secondary-school science also led to an increase in actual performance among the students expected to have the lowest performance. That's likely to in turn have led to an increase in the sense of competency measured in these surveys, consistent with the arrow direction chosen in ref. (*15*) but opposite to that in ref. (*16*). For prolonged traits both directions can coexist.

d) Admissions Criteria

A recent paper examined the causes of admissions to selective graduate programs in physics, focusing on the roles of Graduate Record Exams in physics (GRE-P) and undergraduate grade point averages (GPA), employing an extensive SEM analysis of the causal role of these variables and others on admissions. (*10*) The causal theme is to see what effects dropping GRE-P would have on admissions, and in particular whether GRE-P is helping some students be accepted who might otherwise be overlooked. I shall discuss here only problems in the translation of statistical claims to causal claims, not those in the conversion of data to statistical claims. These problems include incorrect use of SEMs, particularly with regard to causal mediators, and unrealistic framing of the potential outcomes for the interventions under consideration.

The SEM analysis emphasized testing for mediation (one variable serving as a link on the casual path between two others) and moderation (one variable changing the causal effect of another on a third). (*10*) Although the verbal definition of causal mediation is presented correctly, the mathematical method used treats a variable (e.g. GRE-P) as mediating another cause (e.g. GPA) whenever the "mediator" can be expressed in terms of the other cause plus a random component, as explicitly shown in their Equation 2. (*10*) The linear SEM analysis concluded that GRE-P "mediates" the effect of GPA on admissions because the dependence on GPA of the logit for the admissions probability changes when GRE-P is held constant. (*10*)

Nevertheless, it is evident that the choice of which variable to call a mediator is arbitrary, since the relation is essentially symmetrical: the least-squares estimate of each normalized variable is simply the correlation coefficient r times the other normalized variable, plus a variable that's orthogonal to the predictor variable and has variance $(1-r^2)$. This familiar symmetry is just the simplest case of the general graphical equivalence we have seen before. (*1, 14*) Following the procedure of (*10*) would produce evidence of "mediation" in all cases where two causal variables are correlated, regardless of whether any actual causal mediation exists.

No discussion is included as to why relations such as GPA→GRE-P are assumed. (*10*)   If the variables are understood as proxies for underlying traits the relation GRE→ GPA seems about equally plausible, in which case GPA would mediate the effects of GRE-P on admission, rather than the reverse. A more conventional view might be that a variety of more general causes, e.g. diligence in studying physics, affect both variables. That more plausible pattern would be represented by a bidirectional arrow.

The mistaken mathematical conception of mediation leads to at least one interpretive slip that would affect policy choices. The SEM on the effects of "Race" (Fig. 13 of (*10*), reproduced here as Fig. 3) is interpreted to show that GRE-P mediates the causal effects of GPA on admission, including the effects of "Race" on GPA. What the correlations themselves show, however, is that "Race" has very little correlation with GRE-P (not statistically significant in the sample) when GPA is held constant. In a very plausible model in which the correlation between GPA and GRE-P arises mostly from mutual dependence on unmeasured variables rather than a direct causal effect, the standardized URM→GPA coefficient remains more negative than the standardized URM→GPA coefficient.  That means that replacing GRE-P with more weight on GPA could plausibly have decreased the admission of under-represented minorities in this cohort, a conclusion that matters for the policy choices under consideration. For gender, in contrast, such a substitution would favor females, as can be seen from the coefficients in Fig. 12 of (*10*).

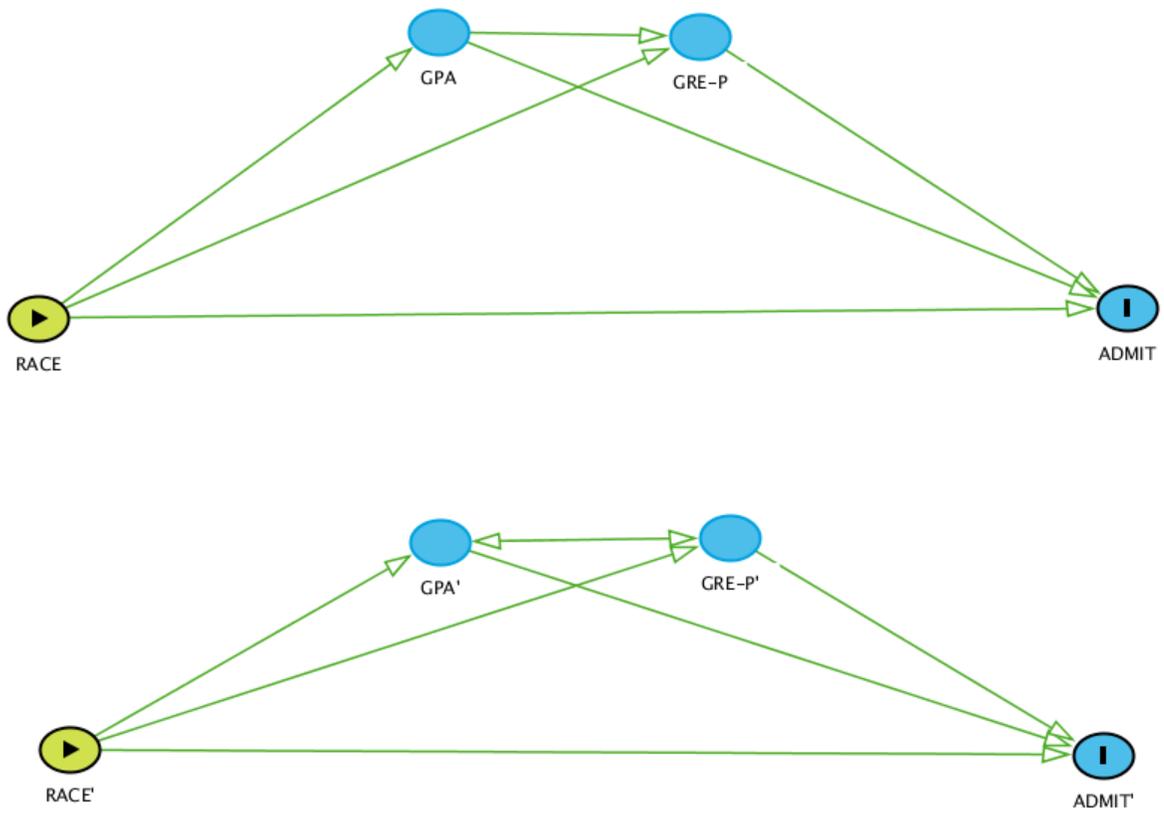

Fig. 3. The SEM graph used to describe the causal effect of "race" on admissions, as given in (*10*) on top, and using a Markov equivalent and seemingly more plausible ADMG on the bottom.

Another fundamental issue arises in the framing of the causal effects, i.e. potential outcomes, of changing admissions criteria. The paper emphasizes that "more applicants could be penalized for having a low physics GRE score despite a high GPA than could benefit from having a high physics GRE score despite a low GPA." (*10*) This framing does not consider the actual potential outcomes, in which the total number of accepted applicants is not set by admission criteria but by limits on funding, mentorship, and job openings. Any change of admission criteria may be judged on a variety of grounds, but these do not include effects on the total number of accepted applicants, which can only be changed by other methods. The comparison of net acceptance rates for different criteria may, however, make sense if the implicit goal is to increase the number of domestic students by reducing the number of international students, who were omitted from the analysis. (*10*)

**Causation within and outside a graph**

The papers discussed above do not adequately consider the variety of different causal relations that might exist among the measured traits. I now turn to a paper that uses SEM analysis properly for the traits measured, but then draws strong causal conclusions about unmeasured causes not explored in the SEM analysis. (*27*) I will illustrate the inability of the data to support the causal claims by showing a DAG including those unmeasured causes.

The paper(*27*) claims that the differences between major demographic groups in introductory college physics test scores are due to differences in pre-course "preparation". The explicit data analysis given shows that when college exam scores of individuals are predicted from a combination of pre-course scores on math ACT/SAT tests and the physics "concept inventory" (CI) exams, then the coefficients of demographic variables included in the predictive model are not large enough to consider statistically significant in this sample, with the possible exception of "first generation" status. The point estimates of these demographic coefficients are small for practical purposes.(*27*) That shows that the incoming test combination (unlike either incoming test on its own) is an approximately unbiased predictor of college physics exam scores with respect to the demographic variables considered. That result could be useful in evaluating interventions designed to reduce demographic disparities, helping in making comparisons of outcomes in non-randomized comparison groups.

On the way from this valid result toward policy suggestions, however, the paper makes a clear causal claim about what the effects of an action would be: "This work shows that creating instruction that enhances the success of every student across the full range of incoming preparations is also the solution to eliminating gaps in the performance across demographic groups." (*27*) That conclusion sounds reasonable and may be supported by other work, but this paper itself has no data on different teaching methods, so it can't show what their effects would be on different students.

The main result was interpreted to mean that "preparation gaps" are responsible for the demographic differences in college scores. (*27*) "Preparation" is not defined in the article, but other traits such as "test anxiety and stereotype threat" are mentioned as possible alternatives to it, indicating that "preparation" is not meant as a catch-all term for all pre-course traits that might affect the test results. (*27*) If "preparation" is nonetheless meant as a catch-all for all traits that actually reflect facility with the material, then the description of the results in terms of it is somewhat tautological. That does not seem to be the meaning intended in the paper, which describes the pre-course tests as "admittedly crude proxies of incoming preparation".(*27*) Thus the measured "preparation" accounts for only a fraction of the individual pre-course score variance, with the rest of the score variance coming from other causes. This strong meaning, in which preparation is not the only possible cause but is observed to be the only important one for demographic differences, was conveyed without ambiguity in a subsequent editorial entitled "It's Not 'Talent,' it's 'Privilege'" by the senior author, explicitly claiming that reference (*27*) shows that talent plays little role in the individual score differentials.(*28*)

That stronger version of the interpretation is not justified by the data. Any fairly stable individual trait measurable by these exams would give the results observed, so long as its effects on both the pre-course and college tests were fairly large and about the same in each demographic group, and other independent traits were approximately demographically balanced. Talent, discussed but rejected in the subsequent editorial(*28*), or interest, discussed in the papers described above(*15, 16*) are familiar traits often considered to affect physics learning but that would not normally be entirely subsumed under "preparation".

A key sentence of the paper comes in the Discussion: "We initially expected that it would be differences in what high school physics courses were taken, but we analyzed that for HSWC [the highly selective west-coast university] , and we found that all demographic groups at this institution had the same distribution of taking AP physics, regular high school physics, and no physics, even though the groups had different average CI prescores and math SAT or ACT scores." (*27*) Thus course work, the main conventional component of preparation, was at least roughly measured in one college and was nominally the same for the different groups. That might suggest that preparation is *not* the key causal variable that differs. Other factors for which the pre-course tests also serve as crude proxies have no such measured indicators of being matched between the different groups.

Realistically, however, courses with the same name can be radically different in different U.S. schools, and those differences are likely to show major correlations with racial/ethnic differences. Therefore, although the results provide no evidence for the role of preparation as distinguished from other causes, they also do not provide evidence against its role in causing differences between those groups. For the most part, however, males and females go to the same schools, so that if they took the same nominal courses they took the same actual courses. Thus the results provide evidence that it is not the most obvious element of preparation, course work, that accounts for the male/female differences in both pre-course and college exams. If "preparation" is taken in a broader sense to include all sorts of pre-test factors, including unequal treatment of the genders both in and out of class, it becomes impossible to distinguish from other causes using these sorts of data.

These relations may again be easier to see with the help of a diagram. The paper (*27*) includes several DAGs used for SEM calculations of the predictive role of math ACTs and CI tests for different demographic traits. Although the paper acknowledges that such SEM analysis "does not test for causality", the actual SEMs used can reasonably support a causal interpretation since the time order is clear: demography first, then pre-tests, lastly course tests. (*27*) Fig. 4 here shows a simplified version of that diagram, but expanded to include the possible causes about which the paper makes claims. For simplicity I include only gender as a demographic variable and do not disaggregate the pre-tests, whose details are irrelevant here and were described in the original paper.(*27*) The problem is not with inference within the part of the graph shown in the paper, but rather that the key causal conclusions concerned potential causes that are not included in that part.

The central observational result is that the sum of the contributions of any paths from gender to post-scores that bypass pre-scores is negligible. That result is unrelated to the causal claim about the differential effects of hypothetical changes in teaching methods(*27*), for which there are no relevant variables in the graph or the data. The causal claims about preparation concern the paths from gender to pre-scores, another region of the graph for which no relevant data are given. The only reported mediator on that path is high-school coursework, but since the effect of gender on it within this limited sample is said to be about zero(*27*), gender differences give no information on the effect of this major type of preparation on downstream variables. Thus the data were consistent with erasing the arrow from Gender to HS Courses, if we ignore their effects on entry into this select cohort. All the other features are unmeasured. These include direct effects of gender on scores, and effects mediated by other forms of preparation or other experiences. All of these are moderated by the effects of the gender-asymmetric social environment, including biased treatment. Inclusion of the pre-test variable tells us that some of these unmeasured variables have fairly long-lasting effects, but does not identify which variables.

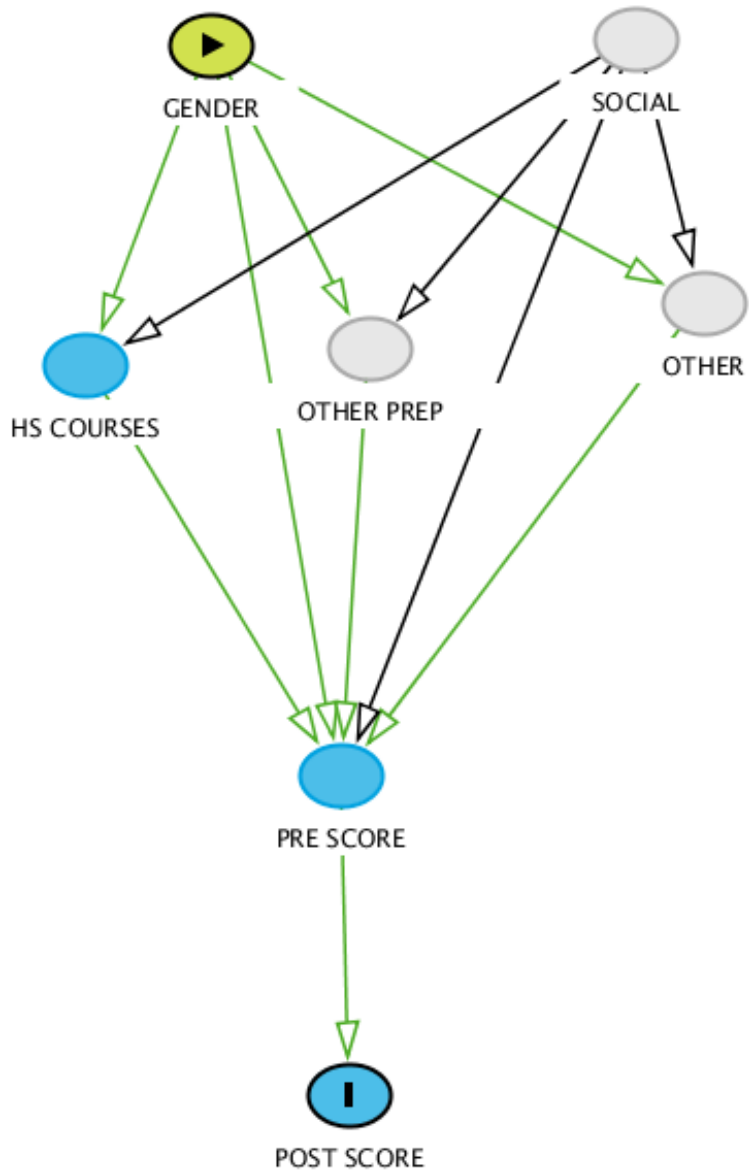

Figure 4. A DAG schematically representing the causal issues addressed in (*27*). Variables shown by lightly-filled ovals are unmeasured. The combined effects of social variables and gender are presumably strongly non-linear, i.e. the effects of gender depend on social context including bias. Since it happens that in this sample the variable HS Courses is approximately independent of Gender(*27*), the arrow between them could be erased. The observational result reported is that no arrow is needed from Gender to PostScore. Other demographic variables have similar diagrams but with poorly measured HS Courses.

The paper offers some options for specific policy recommendations. One broad option is to "provide more resources", (*27*) which presumably would help regardless of the specific causes of the individual differences and is not based on the study's data. Another option seems quite compatible with their interpretation of their data: "adding courses to the sequence to provide a greater range of students the opportunity to start with a course matched to their preparation." (*27*) That policy has long since been implemented in some universities, e.g. my own, and it would be valuable to have studies of how well it works.

The primary option mentioned, however, is "…adjusting the course level to better match the preparation of the less prepared students…" This option is the one that the senior author amplified in his subsequent editorial based on the paper, advocating "changing the coverage and pace of some intro courses so they are optimized for the third of the distribution with the least preparation…". (*28*)   That policy recommendation seems less consistent with the "preparation" interpretation than with rejected causal interpretations. It may be a good way to modify some physics courses, but the data reported in the paper have no implications for that causal claim.

Choosing Variables to Include in Causal Models

A recent methods paper(*11*) attempts to introduce PRPER readers to some of the issues involved in inferring causation from correlations, focusing on bias that can result from omitting variables. Although some of the verbal introduction is correct, the rules are stated incorrectly and the examples used undercut the message that one needs to carefully define causal questions before determining the correct combination of correlations needed to infer a causal effect with minimal bias. The title already conveys that bias arises from *omitting* variables from a model, but it is well-known that bias can equally well arise from *including* inappropriate variables in a model.(*1, 2, 22*)

The following general rule is stated(*11*):

> "A variable included in a model will be biased by an omitted variable if the following two conditions are met:
> (1) the correlation between the omitted variable and the included variable is nonzero, and
> (2) the "true" effect of the omitted variable on the dependent variable is nonzero. "

This rule is simply untrue, since it would say that a causal model should include some variables on which the suspected cause and some unmeasured cause of the outcome both have effects, i.e. collide. Including such variables leads to "collider stratification bias", one of the first common errors described in introductory texts. (*1, 22*) Even more simply, the rule would lead to exclusion of all causal paths that pass through measured mediating variables, allowing one to show that essentially any causal effect is zero by including enough mediators in the model. A paradigmatic case would concern smoking and lung cancer. One could control for tar deposits in the lungs, finding that smoking itself has little effect on lung cancer- instead it's the tar. That gives the wrong answer to the actual question we care about : does smoking itself make someone more likely to get lung cancer? The answer is yes because smoking causes the tar that causes the cancer.

The introduction (*11*) promises to use "explanatory" models for "testing causal hypotheses" in some examples. Nevertheless, the examples (*11*) consistently contrast different "predictive" models without specifying causal patterns. As a result, coefficients in some models are described as "biased" as compared to coefficients in others. This reflects a fundamental misunderstanding of the meaning of "bias". For any set of predictive variables, a properly fitted model gives the true (no quotes) best prediction based on that set of variables. Bias, on the other hand, refers to the systematic mis-estimation of a parameter, in this context a causal parameter. The different coefficients found in different predictive models employing different predictive variables are not "biased"; they are simply the coefficients of the different models. They become biased only if they are used to estimate something else- the causal effects. It is meaningless to say which coefficient is "true" and which is "biased" without specifying the true causal model.

The paper goes on to discuss various cases in which adding a variable to a multiple regression model makes either large or small changes in the coefficients of other variables. Throughout, the assumption is made that the extra variables should always be used when they make large changes, in order to approximate the "true" model (quotes in original). The question investigated is "whether omitting a particular variable will lead to bias…" but the other half of the question, whether including a variable will lead to bias, is simply not mentioned. As we have seen, this approach is fundamentally mistaken because inclusion of covariates (e.g. colliders or mediators) often systematically biases the relevant model coefficient away from the desired causal coefficient.(*1-8*).

The first example given (*11*) follows the same logical pattern as the smoking→tar→ cancer example. It concerns improved outcomes for males, compared to females, when labs are switched from real-world to virtual reality (VR). The causal question is not explicitly defined, but in the context it appears to be finding the difference of VR_benefit for males vs. VR_benefit for females. (*11*) A raw value can be obtained from a simple comparison. The paper claims, however, that one should include another variable, video game experience, which correlates very strongly with gender and strongly with the treatment benefit. (*11*) A first glance at the simple causal diagrams shows that including this variable is inappropriate, since it eliminates a causal path from male→VR_benefit, biasing the net estimate of the male→VR_benefit causal effect. This provides a standard example of how inclusion of a mediator in a model can bias the causal estimate.

We can again use diagrams to help clarify the picture. Fig. 5 shows two diagrams that could account for the correlation between gender and VR_benefit. The top way is the common-sensical one, in which being male leads, in the particular social context, to video gaming. Being male can lead to VR_benefit by the video game path or by other paths. Because the correlation between maleness and video gaming is said to be very high(*11*), collinearity may make it hard to disentangle how much of the male→VR_benefit is mediated by games and how much follows other paths. Nevertheless, the causal effect of being male on VR_benefit is just given by the unconditional regression coefficient in this simple model. On the other hand, inclusion of gaming as an independent predictor *biases* the estimate of the causal effect of maleness on VR_benefit because it eliminates one of the causal paths for that effect, as illustrated in Fig. 5. The paper suggests "running the analysis with video game experience in lieu of gender", which corresponds to erasing not only the arrow from gender to gaming but also the arrow from gender to VR_benefit, in effect treating gender as an irrelevant marker for the gaming.

For that example, since the treatment (being male or female) is not one on which pre-natal intervention is common in our society, giving incorrect causal coefficients for its effects has little direct practical import. The method used, however, would consistently give incorrect causal coefficients for treatments on which interventions are common, such as different educational policies. For a public policy example, if an (unfortunately not fictional) proposed treatment is reduction of number of ballot drop boxes to one per county and the outcome of interest is racial disparity in voting rate, controlling for county population (correlated with the box reduction treatment and having a causal effect on the vote likelihood) would disguise the disparate racial effect.

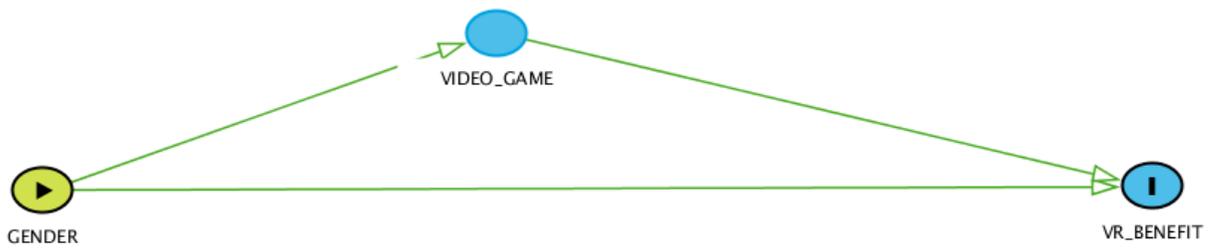

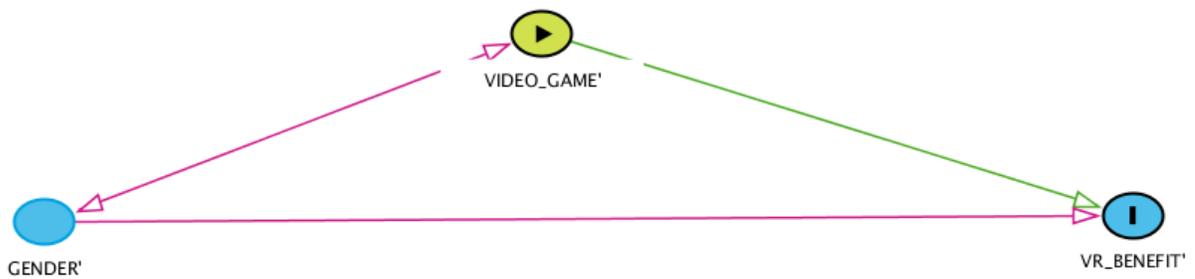

Fig. 5. The top figure represents a common-sense view of a causal relation between gender, gaming, and response to virtual-reality labs. Gender is an exogenous variable, i.e. its causes are not shared with other variables because they are almost always a quasi-random conception lottery. The second graph is the one implicitly used to obtain the Gender→ VR_Benefit coefficient in which gaming and gender are treated as correlated variables neither of which causes the other. One interpretation given by (*11*) would also drop the arrow from gender to VR_benefit.

**Discussion**

Ideally policy recommendations should be based on reliable estimates of consequences. When the recommendations are instead intuitive guesses unconnected to our data, we should be open about that. The difficulties in drawing reliable causal conclusions from observational data in fields such as education should be taken as a reason to pay even closer attention to the relation between models and data than we do in physics, rather than as a reason to relax that attention. That task requires familiarity with modern causal inference methods.

We have seen that several papers ignore the variety of causal possibilities consistent with a given data set. Conclusions are drawn about causal effects without any support in the data. The problem includes two familiar issues in causal inference: that correlations do not give the direction of causation and that they can be altered either by omitting confounding variables from a model or by including ones that themselves bias the causal estimand. (*1*)(*2*) We've also seen policy recommendations that are at best loosely connected with the causal interpretations. Two papers in particular, one with an extensive methods section on mediation(*10*) and the other devoted to describing methods to choose variables(*11*), explicitly advocate methods that are flatly wrong in consequential ways, suggesting a systemic weakness in the field rather than just routine occasional scientific slips.

The papers discussed here were chosen only to illustrate some of the pitfalls on the path from data to policy via causal inference. That it was not hard to find such papers from leading authors in a leading journal indicates that there is room for improvement in the editorial process. Adding some experts in causal inference to the editorial board of PRPER (and other journals) might be helpful. One colleague suggested that perhaps a checkbox for referees to indicate whether their reviews included evaluating causal inference would alert editors to the need for that further step without sacrificing the current evaluations of other aspects of the papers.

It might also help to have a causal reasoning primer specifically for PER, similar to those mentioned for other areas (*3-9*). Its authors should include at least one with domain-specific PER knowledge and one with solid grounding in causal inference, both of which I lack. One referee suggested that a "healthy conversation" on causality would be good for the field, and there are a variety of organized methods by which such a conversation might be held, just as we use for other scientific research and education. I think the most important prerequisite for any such conversation would be the participation of true causal inference experts from outside the PER community.

Since I have suggested possible treatments it's important to also suggest how their effects might be evaluated. If an editorial change is made abruptly, perhaps a regression discontinuity analysis(*29*) could indicate whether it succeeded for the journal, at least if some reliable outcome measure were found. It would then be interesting to look for violations of the Stable Unit Treatment Value Assumption(*30*), i.e. whether the level of causal reasoning was changed in just one journal, or perhaps raised throughout the field, or whether papers with problematic causal reasoning just moved to different journals. The effects of a primer or of some organized conversations would probably be more gradual and thus somewhat harder to disentangle from confounders.

In the meantime, readers who wish to do quick checks of whether reported observations are likely to be causal might start with the traditional informal criteria developed by Bradford Hill for public health questions.(*31*) Perhaps the most compact yet technically sophisticated modern account relevant for PER practitioners is also from the public health field(*4*), which shares many issues with educational research(*9*). As a reader, I've found it useful to sketch a causal graph and interpret words in a paper as statements about links that are on the graph or that are omitted. A fuller description of useful methods and of the many pitfalls (reverse causation, confounding variables, improper controls eliminating causal paths, various forms of selection bias, lack of generalizability, etc.) can be obtained from many texts, e.g. refs. (*1*) and (*2*). That is far beyond the scope of this brief paper, which is intended only to call attention to the issues.

**Acknowledgements**

I thank Carl Wieman for a cordial exchange and Jamie Robins, Sander Greenland, and Thomas Richardson for very helpful editorial comments on sections of this paper and guidance on some very basic graph algebra. I thank the PRPER editors and referees for suggesting this project and for major guidance in revisions.